\documentclass{emulateapj}
\usepackage{natbib}
\shorttitle{Chemistry in an Evolving Protoplanetary Disk}
\usepackage{color}
\newcommand\red{\textcolor{black}}
\begin{document}
\title{Chemistry in an Evolving Protoplanetary Disk: Effects on Terrestrial Planet Composition }
\author{John Moriarty}
\affil{Department of Astronomy, Yale University, New Haven, CT 06511, USA}
\email{john.c.moriarty@yale.edu}
\author{Nikku Madhusudhan}
\affil{Departments of Physics and Astronomy, Yale University, New Haven, CT 06511, USA}
\affil{Institute of Astronomy, University of Cambridge, Madingley Road, Cambridge CB3 0HA, UK}
\and
\author{Debra Fischer}
\affil{Department of Astronomy, Yale University, New Haven, CT 06511, USA}

\begin{abstract}

The composition of planets is largely determined by the chemical and dynamical evolution of the disk during planetesimal formation and growth. To predict the diversity of exoplanet compositions, previous works modeled planetesimal composition as the equilibrium chemical composition of a protoplanetary disk at a single time. However, planetesimals form over an extended period of time, during which, elements sequentially condense out of the gas as the disk cools and are accreted onto planetesimals. To account for the evolution of the disk during planetesimal formation, we couple models of disk chemistry and dynamics with a prescription for planetesimal formation. We then follow the growth of these planetesimals into terrestrial planets with N-body simulations of late stage planet formation to evaluate the effect of sequential condensation on the bulk composition of planets. We find that our model produces results similar to those of earlier models for disks with C/O ratios close to the solar value (0.54). However, in disks with C/O ratios greater than 0.8, carbon rich planetesimals form throughout a much larger radial range of the disk. Furthermore, our model produces carbon rich planetesimals in disks with C/O ratios as low as $\sim$0.65, which is not possible in the static equilibrium chemistry case. These results suggest that (1) there may be a large population of short period carbon rich planets around moderately carbon enhanced stars (0.65 $<$ C/O $<$ 0.8) and (2) carbon rich planets can form throughout the terrestrial planet region around carbon rich stars (C/O $>$ 0.8).

\end{abstract}

%%%%%%%%%%%%%%%%%%%%%%%%%%%%%%%%%%%%%%%%%%%%%%%%%%%%%%%%
\section{Introduction}
With the detection of over 700 confirmed exoplanets and over 3,000 exoplanet candidates \citep[exoplanets.org;][]{Wright11} in a wide variety of planetary system architectures, it has become abundantly clear that there are a diversity of outcomes to the planet formation process. Furthermore, planet mass and radius data coupled with interior structure models, as well as transmission and emission spectroscopy, indicate that exoplanets may be as diverse in their bulk compositions as they are in their orbital properties.  Of the small but growing sample of exoplanets for which mass and radius and/or spectroscopic data exist, a range of bulk compositions have already been inferred, including: Mercury-like compositions (e.g. CoRoT-7b and Kepler-10b) with iron cores composing as much as 65 wt-\% of the planet \citep{Wagner12} to water-worlds  \citep[e.g. GJ1214b;][]{Berta12}. There are even indications of planets with carbon rich atmospheres \citep[e.g. WASP-12b;][]{Madhusudhan10} and interiors \citep[55 Cancri e;][]{Madhusudhan12}.

The potential diversity of terrestrial exoplanet compositions has been addressed from a theoretical standpoint in several publications \citep{Kuchner05, Bond10b, Carter-Bond12, Carter-Bond12b}. The most recent three of these works sought to predict the range of exoplanet compositions that should exist based on the range of elemental abundances that have been observed in planet hosting stars. It was assumed that the stellar abundances reflect those of the initial protoplanetary disk and consequently there should be a connection between the composition of planets that formed from this disk and their host star. The simulations consisted of models of protoplanetary disk chemistry and late stage planet formation, which were coupled together by tagging the initial planetesimals of the planet formation simulation with the composition of the disk's solid material at their initial location. The planets formed in these simulations had compositions ranging from Earth-like to almost entirely C and SiC. These carbon rich planets form in disks with C/O $>$ 0.8, where carbon readily condenses to form solids \citep{Larimer75}.

The composition of the disk's solid material in these simulations was determined with equilibrium chemistry calculations assuming the mid-plane pressure and temperature profile of the disk. Because the mid-plane pressure and temperature profiles of a disk change as the disk ages, it is not obvious which disk age to use when calculating equilibrium abundances. In their work, they adopted an age from a prior study \citep{Bond10} that best reproduced the composition of the Solar system's terrestrial planets.
\citet{Elser12} attempted a more self-consistent approach to determining the disk age for the equilibrium chemistry calculation by imposing a transition condition: the surface density of solid material predicted by the disk model must match the initial surface density of the dynamical simulations. With this condition they had difficulty reproducing the abundances of the Solar system terrestrial planets and found that their results varied depending on the disk model used.

In both of these approaches, the composition of the solids in the disk are determined assuming they all form at the same time. That is, equilibrium composition is calculated at a single age in the disk's lifetime and all solids are assigned that composition. Age dating of meteoritic material suggests that solid material condensed out of the Solar Nebula over the course of about 2.5 Myr or more \citep{Amelin02}. During this time, the Solar Nebula would have cooled significantly resulting in solids being formed over a range of temperatures. 

As a protoplanetary disk cools, it will sequentially condense out each of the elements starting with the most refractory and progressing towards the most volatile. At the same time, the disk is also losing mass. Elements that condense at later times/cooler temperatures will be less abundant and therefore depleted compared to the more refractory elements. \citet{Cassen96, Cassen01} showed that taking into account the cooling of the Solar Nebula during planetesimal formation led to moderately volatile element ratios that were consistent with meteoritic abundances. \red{\citet{Ciesla08}, re-examined this result, this time incorporating the inward migration of small planetesimals, and found that the moderately volatile depletion patterns in the asteroid belt could only be produced in models with a narrow range of parameters that were inconsistent with the ~3 million year timescale of planetesimal formation. They did, however, find that such depletion patterns could occur closer to the Sun, potentially affecting the composition of some of the terrestrial planets.} 

In this paper, we take an approach similar to \citet{Cassen96} and couple models of protoplanetary disk evolution, equilibrium chemistry and planetesimal growth in order to predict the chemical composition of planetesimals and the planets that they form.  We apply this method to systems with solar-like compositions and systems with high C/O ratios and show that considering the growth of planetesimals in an evolving disk affects the chemical composition of the resulting planets. Lastly, we consider the effect that the depletion of oxygen in the outer disk has on planetesimal compositions in the inner disk.

%%%%%%%%%%%%%%%%%%%%%%%%%%%%%%%%%%%%%%%%%%%%%%%%%%%%%%%%%%%%%%

\section{Methods}

We simulate the formation of terrestrial planets with a coupled chemical and dynamical model. A theoretical model of the protoplanetary disk provides the radial temperature, pressure and density structure of the disk as a function of time. We assume that the elemental abundances of the host star reflect the original elemental abundances of the disk. Combining the disk model and the stellar abundances, we calculate the chemical equilibrium composition of the disk as a function of radius. This provides a chemical model of the disk over time which is combined with a prescription for the planetesimal formation rate to determine the composition of planetesimals that form in the disk. We then follow the dynamical evolution of a population of planetesimals as they collide and build planets. The final composition of these planets is the combination of each of the planetesimals that they accreted.

	\subsection{Disk Model}
	\label{diskmodel}
	The pressure and temperature structure of the disk, necessary for the equilibrium chemistry calculations, is calculated using a theoretical model derived in \citet{Chambers09} which describes a viscously heated and irradiated disk.

	The Chambers model divides the disk into three regions. In the intermediate region the heating of the disk is dominated by viscous dissipation. The surface density is given by
	\begin{equation}
	\Sigma(r,t) = \Sigma_{vis}\left(\frac{r}{s_0}\right)^{-3/5}\left(1+\frac{t}{\tau_{vis}}\right)^{-57/80},
	\end{equation}
	where
	\begin{equation}
	\Sigma_{vis}=\frac{7M_0}{10\pi s_0^2}, 
	\end{equation}

	and $M_0$ and $s_0$ are the initial mass and outer edge of the disk.
	The temperature of the disk in this region is given by
	\begin{equation}
	T(r,t)=T_{vis}\left(\frac{r}{s_0}\right)^{-9/10}\left(1+\frac{t}{\tau_{vis}}\right)^{-19/40},
	\end{equation} 
	where
	\begin{equation}
	T_{vis}=\left(\frac{27\kappa_0}{64\sigma}\right)^{1/3} \left(\frac{\alpha\gamma k_B}{\mu m_H}\right)^{1/3} \left(\frac{7M_0}{10\pi s_0^2}\right)^{2/3}  \left(\frac{GM_*}{s_0^2}\right)^{1/6}. 
	\end{equation}
	$\kappa_0$ is the opacity and is taken to be a constant 3 $cm^2g^{-1}$, $\alpha=0.01$ is the viscosity parameter, $\mu=2.4$ is the mean molecular weight, $\gamma=1.7$ is the adiabatic index, and $m_H$ is the mass of hydrogen. The viscous timescale is:
		\begin{equation}
	\tau_{vis} = \frac{1}{16\pi}\frac{\mu m_H}{\alpha\gamma k_B}\frac{\Omega_0M_0}{\Sigma_{vis}T_{vis}},
	\end{equation}
	
	In the inner region of the disk, heating is still dominated by viscous dissipation but the temperature is too high for solids to condense and consequently the opacity is expected to be several orders of magnitude lower than elsewhere in the disk. The opacity is assumed to follow the power law form given in \citet{Stepinski98},
	\begin{equation}
	\kappa=\kappa_0\left(\frac{T}{T_{e}}\right)^n,
	\end{equation}
	where $T_e=1380K$ is the temperature at which the majority of solids evaporate, $\kappa_0$ is the opacity at $T<T_e$ and $n$ is the power law index taken to be -14.
	The surface density in the inner region of the disk is,
	\begin{equation}
	\Sigma(r,t) = \Sigma_{evap}\left(\frac{r}{s_0}\right)^{-24/19}\left(1+\frac{t}{\tau_{vis}}\right)^{-17/16},
	\end{equation}
	where
	\begin{equation}
	\Sigma_{evap} = \Sigma_{vis}\left(\frac{T_{vis}}{T_e}\right)^{14/19}. 
	\end{equation}
	The temperature in the inner region is given by 
	\begin{equation}
	T(r,t)=T_{vis}^{5/19}T_{e}^{14/19}\left(\frac{r}{s_0}\right)^{-9/38}\left(1+\frac{t}{\tau_{vis}}\right)^{-1/8}.
	\end{equation}
	The transition radius between these regions is
	\begin{equation}
	r_e(t) = s_0\left(\frac{\Sigma_{evap}}{\Sigma_{vis}}\right)^{95/63}\left(1+\frac{t}{\tau_{vis}}\right)^{-19/36}.
	\end{equation}
	
	In the outer regions of the disk, viscous dissipation does not deposit significant amounts of energy and so stellar irradiation dominates the heating of the disk. The surface density equation is 
\begin{equation}
	\Sigma(r,t) = \Sigma_{rad}\left(\frac{r}{s_0}\right)^{-15/14}\left(1+\frac{t}{\tau_{vis}}\right)^{-19/16},
\end{equation}
	where,
	\begin{equation}
	\Sigma_{rad}=\Sigma_{vis}\left(\frac{T_{vis}}{T_{rad}}\right),
	\end{equation}
	and
	\begin{equation}
	T_{rad}=\left(\frac{4}{7}\right)^{1/4}\left(\frac{T_*kR_*}{GM_*\mu m_H}\right)^{1/7}\left(\frac{R_*}{s_0}\right)^{3/7}T_*.
	\end{equation}
	The temperature in the outer region is 
	\begin{equation}
	T(r,t)=T_{rad}\left(\frac{r}{s_0}\right)^{-3/7}.
	\end{equation}
	The transition radius between the outer irradiated region and the intermediate viscous region is given by
	\begin{equation}
		r_t(t) = s_0\left(\frac{\Sigma_{rad}}{\Sigma_{vis}}\right)^{70/33}\left(1+\frac{t}{\tau_{vis}}\right)^{-133/132}.
	\end{equation}
	
	For all calculations in this work we use $M_0=0.1M_{\odot}$, $s_0=33AU$, $T_*=4200K$ and $R_*=3R_{\odot}$.  This model corresponds to the second example in \citet{Stepinski98} which is consistent with a planetesimal forming disk.
	 
	\subsection{Equilibrium Chemistry}
	Earlier works \citep[e.g.][]{Bond10,Elser12}, have used equilibrium chemistry calculations in the protoplanetary disk to predict planetesimal compositions. Each calculation used the pressure and temperature profile of a protoplanetary disk at one point in its evolution and so the composition of the planetesimals reflect the composition of the disk only at that epoch. The planets built from these planetesimals had compositions that, to first order, were consistent with the bulk compositions of the terrestrial planets of the Solar System. As a comparison with these works, we perform similar equilibrium calculations.
	
	 Like the aforementioned works, we use the software package HSC Chemistry (version 7.1). Given the pressure, temperature and elemental abundances, HSC Chemistry will calculate the equilibrium abundances for a list of species by minimizing the total Gibbs free energy of the system using the GIBBS solver \citep{White85}. The species considered are the same as those used in \citet{Bond10} and are listed in Table \ref{Species}. Each of the solid species was considered a pure substance (i.e. no solid solutions). Calculations were performed between 0.3 and 4 AU in radial extent to match the region considered in the dynamical simulations.

\red{ The accuracy of our chemical model is limited by two main factors: 1) The completeness of the list of chemical species considered. This list is limited by both the computational complexity of calculating equilibrium for a large number of species and HSC Chemistry's database. A noticeable shortfall is the lack of carbon in our models throughout the asteroid belt (Figure~\ref{DiskComposition}) where carbonaceous meteorites can contain as much as few percent by mass in carbon. This is expected, though, because most carbon in carbonaceous meteorites is in the form of organic macromolecules \citep{Pizzarello06} which are not included in our study. 2) The assumption of equilibrium conditions. It is unknown exactly how much of the Solar Nebula experienced conditions where chemical equilibrium could occur. However, the material that makes up the asteroids and terrestrial planets of the Solar System is believed to have formed under near equilibrium conditions \citep[][and references therein]{Davis07}.
}

\begin{table}
\begin{center}
\caption{Chemical species used in equilibrium calculations.}
\label{Species}
                	\begin{tabular}{llllll}
                	\hline
               	\multicolumn{6}{c}{Gases}\\
	     	\hline
                   Al & CaH & FeS & MgOH & NiS & SiC\\
                   Al$_2$O & CaO & H & MgS & O & SiH\\
                   AlH & CaOH & H$_2$ & N & O$_2$ & SiN\\
                   AlO & CaS & H$_2$O & NH$_3$ & P & SiO\\
                   AlOH & Cr & H$_2$S & NO & PH & SiP\\
                   AlS & CrH & HCN & Na & PN & SiP$_2$\\
                   C & CrN & HCO & Na$_2$ & PO & SiS\\
                   CH$_4$ & CrO & HPO & NaH & PS & Ti\\
                   CN & CrOH & HS & NaO & S & TiN\\
                   CO & CrS & He & NaOH & S$_2$ & TiO\\
                   CO$_2$ & Fe & Mg & Ni & SN & TiO$_2$\\
                   CP & FeH & MgH & NiH & SO & TiS\\
                   CS & FeO & MgN & NiO & SO$_2$ & \\
                   Ca & FeOH & MgO & NiOH & Si & \\

		\hline
                   \multicolumn{6}{c}{Solids}\\
          	\hline
          
            	\multicolumn{2}{l}{Al$_2$O$_3$} &            	\multicolumn{2}{l}{Cr$_2$FeO$_4$} &            	\multicolumn{2}{l}{Mg$_2$SiO$_4$} \\
            	\multicolumn{2}{l}{AlN} &            			\multicolumn{2}{l}{Fe} &            				\multicolumn{2}{l}{Mg$_3$Si$_2$O$_5$(OH)$_4$} \\
            	\multicolumn{2}{l}{C} &            				\multicolumn{2}{l}{Fe$_2$SiO$_4$} &            	\multicolumn{2}{l}{MgSiO$_3$} \\
              	\multicolumn{2}{l}{Ca$_3$(PO$_4$)$_2$} &  	\multicolumn{2}{l}{Fe$_3$C} &           		\multicolumn{2}{l}{NaAlSi$_3$O$_8$} \\
              	\multicolumn{2}{l}{Ca$_2$Al$_2$SiO$_7$} &  	\multicolumn{2}{l}{Fe$_3$O$_4$} &           	\multicolumn{2}{l}{Ni} \\
              	\multicolumn{2}{l}{CaMgSi$_2$O$_6$} &          \multicolumn{2}{l}{Fe$_3$P} &           		\multicolumn{2}{l}{P} \\
             	\multicolumn{2}{l}{CaAl$_{12}$O$_{19}$} &      \multicolumn{2}{l}{FeS} &           			\multicolumn{2}{l}{Si} \\
             	\multicolumn{2}{l}{CaAl$_2$Si$_2$O$_8$} &  	\multicolumn{2}{l}{FeSiO$_3$} &           		\multicolumn{2}{l}{SiC} \\
             	\multicolumn{2}{l}{CaTiO$_3$} &           		\multicolumn{2}{l}{H$_2$O} &           			\multicolumn{2}{l}{Ti$_2$O$_3$} \\
             	\multicolumn{2}{l}{CaS} &           			\multicolumn{2}{l}{MgAl$_2$O$_4$} &           	\multicolumn{2}{l}{TiC} \\
             	\multicolumn{2}{l}{Cr} &           				\multicolumn{2}{l}{MgS} &           			\multicolumn{2}{l}{TiN} \\

                   \hline
		\end{tabular}
		\end{center}
\end{table}
	 
	\subsection{Sequential Condensation Chemistry}
	
	In a more realistic scenario, planetesimals are formed over the course of a few million years. Because the temperature profile of a protoplanetary disk changes significantly during this period, it is necessary to consider the changes in the composition of solid material available for building planetesimals as the disk ages.
	
\red{Taking a similar approach to \citet{Cassen96}, we start with an initially hot protoplanetary disk and evolve it forward in time while keeping track of the chemical composition of dust, gas and planetesimals in the disk. In our model, we break the disk up into two components: the gas/dust disk and the planetesimal disk. The planetesimal disk is assumed to not react chemically with the gas/dust disk as only the thin outer shell of a planetesimal is in contact with the gas and dust of the disk. The planetesimal disk is initially empty but builds up over time as planetesimals form from the available solid material. To calculate the composition of these planetesimals we iterate through the lifetime of the disk. At each time step and in each radial zone of the disk we perform the following steps:
\begin{enumerate}
     \item Calculate the equilibrium chemical composition of the dust/gas disk using HSC Chemistry.
     \item Remove a fraction of the solid material from the gas/dust disk and add it to the planetesimal disk. The amount of material that is converted to planetesimals is determined by the planetesimal formation rate which is described shortly.
     \item Calculate the radial movement of the gas and dust according to the disk equations described in Section~\ref{diskmodel}. In our model we assume that all dust is perfectly coupled to the gas and that planetesimals are large enough that they do not experience orbital migration due to gas drag and thus remain where they formed.
     \item Recalculate the chemical inventory throughout the disk as it has changed due to radial motion of the disk and formation of planetesimals.
\end{enumerate}
}

\red{We repeat each of these steps until the amount of material being converted into planetesimals is negligible (about 3 million years in our model). The compositions of the planetesimals at the end of this portion of the simulation are taken as the input composition for the next stage of our model (the n-body simulations of late stage planet formation).
}

	As solid material grows from dust to planetesimal size, it must cross a size threshold where it goes from being chemically reactive to mostly locked away in the interior of a planetesimal. We will call the rate at which material crosses this threshold the planetesimal formation rate. The planetesimal formation rate in protoplanetary disks is currently unknown. In fact, the formation of planetesimals, in general, is not well understood. A number of mechanisms have been proposed to explain their formation, which can be categorized into two general processes: 1. Planetesimals formed from the pairwise collisions and sticking of larger and larger aggregates \citep[][and references therin]{Blum08} and 2. They formed from the collective self gravity of a large mass of small particles \citep[e.g][]{Johansen07,Cuzzi08}. 
	
	Regardless of the mechanism of planetesimal formation, the planetesimal formation rate will vary in time and space due to differing conditions in the disk. For the sequential condensation chemistry model it is essential to know how the planetesimal formation rate varies. For lack of a consensus on the planetesimal formation mechanism, we will use the simple prescription for the planetesimal formation rate used in \citet{Cassen96}, which is consistent with formation from coagulation:
	\begin{equation}
	\dot{\Sigma}_p \propto \Sigma_{solid}\Omega,
	\label{planetesimalformationrate}
	\end{equation}
	where $\Sigma_{solid}$ is the surface density of the solid material available to build planetesimals at a point in the disk and $\Omega$ is the orbital angular speed at that point. The exact form of this expression may change as planetesimal formation models are revised. Nevertheless, a planetesimal formation rate that increases with increasing solid surface density and a decreasing dynamical timescale is a logical starting point.

	Despite the fact that only the planetesimals that form interior to 4 AU are considered in further steps of the model (see Section 2.4) it is necessary to track the formation of planetesimals farther out in the disk as well. Planetesimals forming in the outer regions will deplete the dust and gas disk of any elements that are solid, which, when the disk material moves inward, can effect the chemistry in the terrestrial planet forming region. For this reason, our model extends to 10 AU, at which point the maximum depletion of any element is $<$10\%.
	
		As discussed previously, the assumption of equilibrium chemistry is likely valid for the inner regions of the disk. Our model does not take into account deviations from equilibrium chemistry that may occur in the regions of the disk exterior to 4 AU. Because the planetesimals formed in this region are not considered in the dynamical simulations, these deviations are only important to the extent that they change the composition of the inward moving material. Material is depleted no more than 15\% by the time it reaches 4 AU. Therefore, deviations from equilibrium chemistry will only effect less than 15\% of the material.

	\subsection{Dynamical Simulations}
	The final step of our planet formation model is to determine the number and masses of planets that can form in a system and from what radial range they accreted their material. This step is important because it allows us to track the compositional mixing of planetesimals from different regions of the disk. \red{To do this, we tag a population of planetesimals with the final composition of planetesimals in the sequential condensation chemistry model and follow their growth into planets.}
	
	Rather than simulating the entire growth of small planetesimals all the way to planets, we employ the results of earlier works, and begin at the end of the oligarchic growth phase. This allows us to skip a computationally expensive step which is not necessary for our purposes. \citet{Kokubo02} \red{find that accretion through the oligarchic growth phase proceeds locally assuming that the radial migration of small fragments produced in planetesimal collisions does not significantly alter the surface density profile. Consequently, any planetesimal or embryo surviving at the end of this phase should have a composition similar to the original planetesimals at that location.} On the other hand, during the final stage of planet formation, significant radial mixing occurs as planetary embryos scatter and collide with each other and slowly sweep up the remaining planetesimals in the system. 
			
	We begin our late stage planet formation simulations at the end of the oligarchic growth phase. As described in \citet{Kokubo02}, at this point, planetesimals have a bimodel distribution in mass. Planetary embryos are much more massive and make up about 50\% of the mass while the rest of the mass is in smaller planetesimals. The embryos are typically separated by about 10 mutual Hill radii. Following \citet{Chambers01} and \citet{Obrien06}, we start our simulations with such a distribution of planetesimals placed throughout the disk such that the surface density profile of the disk obeys the relation
	\begin{equation}
	\Sigma = \Sigma_0\left(\frac{r}{1AU}\right)^{-3/2},
	\label{surfacedensity}
	\end{equation}
	where $\Sigma_0=8gcm^{-2}$ and then falls off linearly between 0.7 and 0.3 AU. \red{We note that a fully self-consistent model would use the surface density profile calculated from the sequential condensation chemistry model rather than that of Equation 17. However, this would require separate N-body simulations for each different system. Because late stage accretion is a stochastic process, it would then be very difficult to disentangle the effects of the different initial planetesimal compositions from differences in the N-body simulations. As described in Section 3.1 these two surface density profiles are compatible with each other.}
		For these simulations, we use the N-body integrator, Mercury \citep{Chambers99}, which allows for a set of larger bodies that interact gravitationally with each other, as well as a set of smaller bodies that interact gravitationally with the larger bodies but not each other. This is ideally suited for the initial bimodel distribution of planetesimals used. Planets then grow through the collisions of these bodies. The collisions are assumed to be totally inelastic  (i.e. they conserve momentum and form a single body with a mass of the combined mass of the colliding planetesimals). 
		
		We performed a set of four simulations, similar to those of \citet{Obrien06}, with an initial population of 26 embryos of mass 0.09 $M_\oplus$ and about 1000 planetesimals of mass 0.0024 $M_\oplus$ distributed between 0.3 and 4 AU around a solar mass star such that they obeyed the surface density relation given in Equation 17. The eccentricity and inclination of each planetesimal was given a random value between $0-0.01$ and $0-0.5^\circ$ respectively, and the longitude of the ascending node, argument of periapsis and mean anomaly were assigned random values between 0 and 360$^\circ$. Each system was integrated for 250 Myr with a time step of 5 days.

\subsection{System Compositions}
	We chose four systems with stellar abundances that covered the range of Mg/Si and C/O ratios observed in stars. These ratios have the biggest influence on the mineralogy of the disk \citep{Bond10b} and so covering the full range should highlight the major differences between the equilibrium chemistry and sequential condensation models. The stellar abundances we used are the same as those used in \citet{Bond10b}, which were taken from \citet{Ecuvillon04}, \citet{Ecuvillon06}, \citet{Beirao05}, and \citet{Gilli06}, and the solar abundances were taken from \citet{Asplund05}. These abundances are shown in Table \ref{Abundances}. \red{We note that we only use the elemental compositions of these stars in our simulations. We do not use other physical parameters of the systems (e.g., stellar mass and planetary companions) for the N-body simulations. We choose to do this so that the same N-body simulations can be used for each system.}
	
\begin{table}
	\caption{Elemental abundances for systems considered.}
	\begin{center}
      	\begin{tabular}{lllll}
	\hline
   Element & Solar System & 55 Cnc & HD19994 & HD 213240 \\
   \hline
Al  &  2.34$\times10^6$   &  8.7$\times10^6$  &  6.2$\times10^6$  &  4.7$\times10^6$  \\
C  &  2.45$\times10^8$   &  7.4$\times10^8$  &  8.9$\times10^8$  &  5.4$\times10^8$  \\
Ca  &  2.04$\times10^6$   &  2.8$\times10^6$  &  3.4$\times10^6$  &  2.5$\times10^6$  \\
Cr  &  4.37$\times10^5$   &  7.8$\times10^5$  &  7.4$\times10^5$  &  5.2$\times10^5$  \\
Fe  &  2.82$\times10^7$   &  6.3$\times10^7$  &  5.1$\times10^7$  &  4.4$\times10^7$  \\
H  &  1.00$\times10^{12}$   &  1.0$\times10^{12}$  &  1.0$\times10^{12}$  &  1.0$\times10^{12}$  \\
He  &  8.51$\times10^{10}$   &  8.5$\times10^{10}$  &  8.5$\times10^{10}$  &  8.5$\times10^{10}$  \\
Mg  &  3.39$\times10^7$   &  1.2$\times10^8$  &  6.2$\times10^7$  &  6.5$\times10^7$  \\
N  &  6.03$\times10^7$   &  1.8$\times10^8$  &  2.2$\times10^8$  &  1.3$\times10^8$  \\
Na  &  1.48$\times10^6$   &  3.9$\times10^6$  &  6.5$\times10^6$  &  3.6$\times10^6$  \\
Ni  &  1.70$\times10^6$   &  3.6$\times10^6$  &  3.3$\times10^6$  &  2.4$\times10^6$  \\
O  &  4.57$\times10^8$   &  7.4$\times10^8$  &  7.1$\times10^8$  &  1.2$\times10^9$  \\
P  &  2.29$\times10^5$   &  8.5$\times10^5$  &  6.0$\times10^5$  &  4.6$\times10^5$  \\
S  &  1.38$\times10^7$   &  2.1$\times10^7$  &  1.4$\times10^7$  &  1.3$\times10^7$  \\
Si  &  3.24$\times10^7$   &  6.9$\times10^7$  &  6.0$\times10^7$  &  4.4$\times10^7$  \\
Ti  &  7.94$\times10^4$   &  2.2$\times10^5$  &  1.5$\times10^5$  &  1.3$\times10^5$  \\
		\hline

		\end{tabular}
		\end{center}

\label{Abundances}
\end{table}

%%%%%%%%%%%%%%%%%%%%%%%%%%%%%%%%%%%%%%%%%%%%%%%%%%%%%%%%%%%%%%%%

\section{Results}
\subsection{Planetesimal Surface Density}
	Before the sequential condensation model could be used to predict planetesimal composition, it was necessary to calibrate the planetesimal formation rate. Changing the constant of proportionality in the prescription for the planetesimal formation rate will increase or decrease the total mass in planetesimals that have formed by the end of the simulation and will also effect the chemical composition of the planetesimals. The planetesimal surface density that we model needs to match that which we believe existed during the early stages of planet formation. We can place constraints on this planetesimal surface density based on the masses of the terrestrial planets in the Solar System \citep[i.e. the minimum mass Solar Nebula or MMSN;][]{Weidenschilling77}. Once planetesimals are formed, planet formation is a fairly efficient process with planets accreting about 50\% of the original mass \citep{Obrien06}. This means that although the MMSN is a lower limit it is probably not far from the true amount of solid material that existed in the disk. The concept of the MMSN has been applied to extrasolar systems \citep[e.g.][]{Chiang13,Kuchner04} with a range of possible disk masses. In Section 3.2 we consider the effects of changing this normalization factor on our predicted planetesimal compositions.
	
	 In order to match the planetesimal surface density of the MMSN we found that the planetesimal formation rate in our model should be:
\begin{equation}
\dot{\Sigma}_p = 9.5\times10^{-6}yr^{-1}\Sigma_{solid}\frac{\Omega}{\Omega_{1AU}}
\end{equation}
This produced a distribution of planetesimal masses that is consistent with the $r^{-3/2}$ profile of the MMSN. N-body simulations of planet formation in the inner Solar System \citep[e.g.][]{Chambers01, Obrien06} required surface density profiles that decreased inwards of 0.7 AU in order to form low mass Mercury analogs. However, the surface density profile produced in the sequential condensation model continues to increase interior to 0.7 AU. This discrepancy can be explained in a number of ways. For example, the prescription for the planetesimal formation/growth rate may be oversimplified and not appropriate for the innermost region of the disk, or planetesimals may become depleted in this region due to faster migration rates associated with higher gas densities. In our model, when we seed the disk with planetesimals for the dynamical simulations, we do not not assume the surface density profile output by the sequential condensation model. Instead we assume the surface density profile of Equation \ref{surfacedensity}. The choice of surface density profile does not significantly affect the results because they are consistent with each other to begin with. However, we chose to use Equation \ref{surfacedensity} so that we could use the same set of N-body simulations for each of the different system compositions. This allows us to compare the affects of different initial compositions without having to account for the stochastic nature of late stage planet formation.

\subsection{Disk Chemistry}
\begin{figure*}      % use "figure*" instead of "figure" if you want your figure to span both columns
\epsscale{1.3}      % adjust this number to change the size of your figure
\plotone{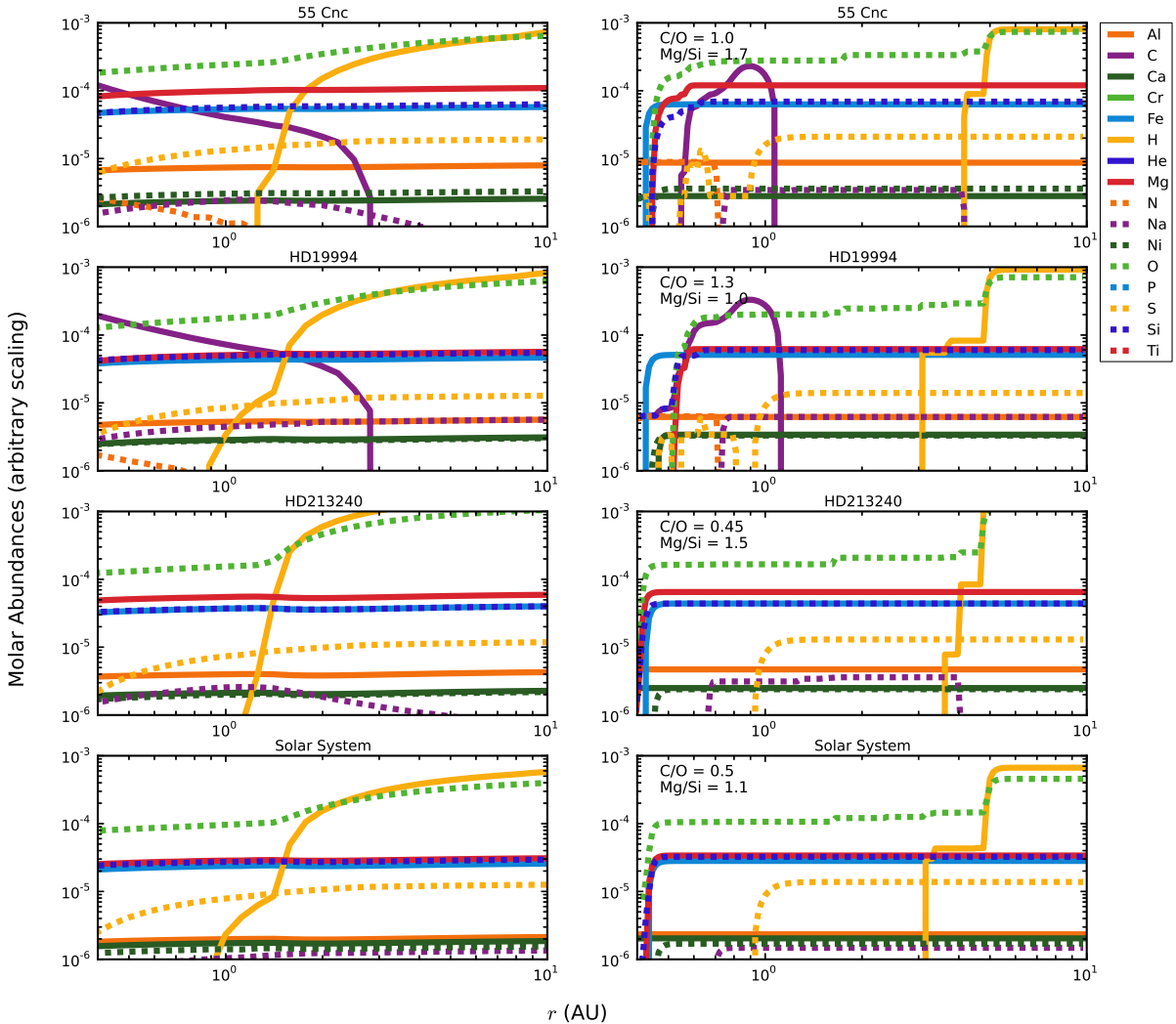}
\caption{Elemental abundances as a function of radius ($r$) in the disk for the four systems considered. The left column shows results from the sequential condensation model and the right column shows results from the equilibrium chemistry model.}
\label{DiskComposition}
\end{figure*}        % again use "figure*" instead of "figure" to span both columns
	Figure \ref{DiskComposition} shows the composition of planetesimals as a function of radius in the disk for each of the four different systems for both the equilibrium chemistry model and the sequential condensation model. \red{In the equilibrium chemistry model, planetesimals were assumed to form instantaneously at $1.5\times10^5$ years. We chose a disk age of $1.5\times10^5$ years because it produces planets with compositions that matched those of the terrestrial planets of the Solar System well.} This figure highlights two major differences that occur by incorporating the time evolution of the disk into the chemistry model: 1. The radial range in which elements are present in planetesimals is extended in the sequential condensation model relative to the equilibrium chemistry model. 2. The sequential condensation model leads to smoothly varying elemental abundances in planetesimals with increasing radius in the disk compared to the equilibrium chemistry model where elements go from not present to fully condensed out over a small radial range.

	The radial range where carbon makes up a significant fraction of the planetesimals' composition is much larger in the sequential condensation model, extending out to $\sim$3 AU, compared to the equilibrium chemistry model where carbon is only present in planetesimals interior to $\sim$1 AU. In carbon rich systems, carbon is present in solid form at temperatures higher than 550 K. In the equilibrium chemistry model, carbon rich planetesimals are limited to the region of the disk where these temperatures occur at the specific disk age used. On the other hand, planetesimals in the sequential condensation model are the accumulation of the solid material formed throughout the lifetime of the disk. This leads to carbon accumulating onto planetesimals farther out in the disk when it was younger and hotter, and closer to the star as the disk cools.
	
	The relative amounts of the most abundant elements in systems with more solar-like chemistry are much less effected by the choice of chemistry model. Figure \ref{DiskComposition}  shows that the ratios of Mg, Fe and Si are relatively unchanged between models for these systems. This is due to the fact that these elements all condense out at about the same temperature (between 1300K and 1330K). The ratio of O to these three elements is also quite similar between the models interior to 2 AU. However, beyond 2 AU the sequential condensation model shows a much quicker increase in O abundance compared to the equilibrium chemistry model. This occurs because the ice line continues to move closer to the star as the disk cools which is not captured in the equilibrium chemistry model. In addition to O, the relative abundances of some of the other more volatile elements can be seen to vary between the two models. Specifically, the amounts of S and Na show a gradual decrease closer to the star in the sequential condensation model compared to the abrupt change seen in the equilibrium chemistry model. \red{Although this gradual decrease is qualitatively consistent with the volatile element fractionation patterns seen in Solar System meteorites \citep[see][for a review]{Palme88}, \citet{Ciesla08} found that this trend cannot be reproduced throughout the asteroid belt where it is observed in the Solar System in sequential condensation-like models. This result suggests that other mechanisms, such as evaporation of volatiles \citep{Davis05}, may play an important role in determining the composition of planetesimals.}
	
	A more subtle effect of the sequential condensation model is that it depletes the disk of certain elements relative to others. As a parcel of gas and dust migrates through the disk, it deposits solid material in the form of planetesimals. The disk becomes depleted in the elements that composed this material, or in a relative sense, enriched in the elements that were not present. The exact amount of depletion seen in any given parcel will depend on where it originated from, the disk model that was used and the planetesimal accretion rate.
	
\begin{figure}      % use "figure*" instead of "figure" if you want your figure to span both columns
\epsscale{1.4}      % adjust this number to change the size of your figure
\plotone{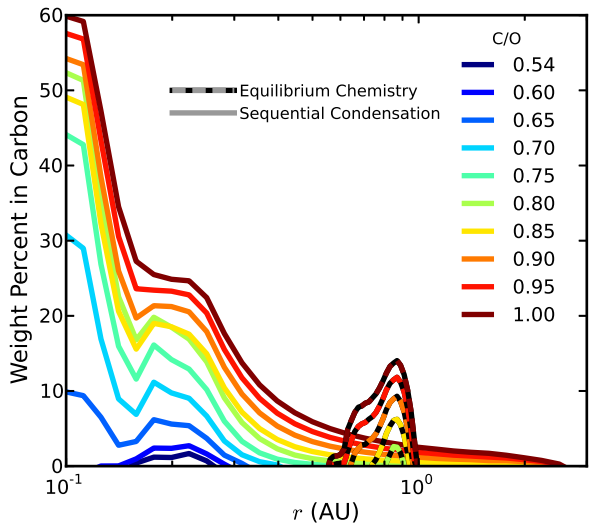}
\caption{The wt-\% of carbon contained in planetesimals as a function of radius ($r$) in the disk. Each line represents a disk with a different initial C/O ratio ranging from 0.54 (solar) to 1. For comparison, the equilibrium chemistry results (dashed lines) are shown in addition to the sequential condensation results (solid lines). The sequential condensation model produces carbon rich planetesimals throughout a larger radial range of the disk and in disks with lower initial C/O ratios than the equilibrium chemistry model does. }
\label{CtoORatio}
\end{figure}

	This effect is extremely important when considering the C/O ratio of a disk. \red{Carbon is abundant as a solid in the form of graphite at temperatures between $\sim$600K and $\sim$1000K, and silicon carbide at temperatures greater than $\sim$1000K. It is also abundant in carbon ices below 78K \citep[][ We do not model these low temperature ices because the disk does not get to low enough temperatures in the region and time of interest.]{Lodders03} For a large portion of the disk (where temperatures are between 78K and 600K), very little carbon will condense out of the disk.} In contrast, oxygen is the dominant component of rocky material (present at temperatures below 1300K) and is even more abundant as a solid beyond the ice line. This means that throughout much of the disk, oxygen is being depleted much more than carbon leading to an increase in the C/O ratio. In some cases the C/O ratio can reach high enough levels that once a parcel moves closer to the star and heats up, carbon will condense out of the disk creating carbon rich planetesimals. This can occur for systems that were not originally significantly enriched in C. Figure \ref{CtoORatio} shows the wt-\% of carbon contained in planetesimals for disks with initial C/O ratios between solar and 1.00. The initial elemental abundances of the disk for each curve are all solar, except for the oxygen abundance which has been decreased in order to obtain each C/O ratio. In contrast to the sequential condensation model, carbon rich planetesimals can only be formed in the equilibrium chemistry model if the initial C/O ratio is greater than 0.8. Furthermore, the C/O enhancement of the sequential condensation model allows a much larger region of the disk to form carbon rich planetesimals.
	
        % again use "figure*" instead of "figure" to span both columns

\begin{figure}      % use "figure*" instead of "figure" if you want your figure to span both columns
\epsscale{1.4}      % adjust this number to change the size of your figure
\plotone{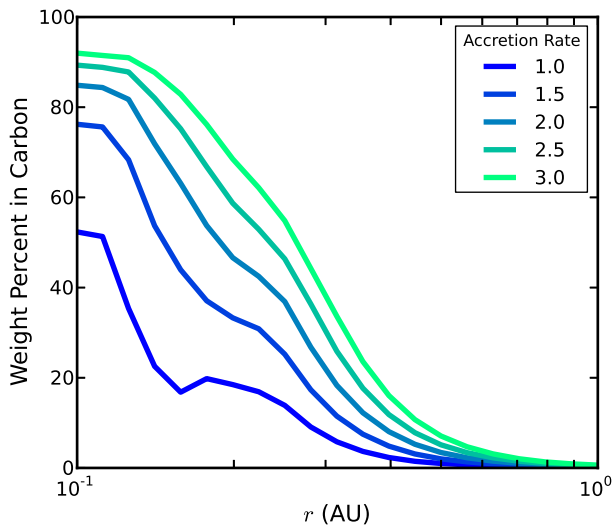}
\caption{The wt-\% of carbon contained in planetesimals at different radii ($r$) in the disk for different planetesimal accretion rates. The planetesimal accretion rate used for each simulation is the nominal value needed to obtain a final planetesimal surface density similar to the MMSN scaled by the value in the legend. }
\label{AccretionRate}
\end{figure}

	The amount of carbon deposited in planetesimals will depend not only on the original C/O ratio of the disk but also on the disk model and planetesimal accretion rate that is used. Exploring the possible range of disk models is beyond the scope of this paper, but their effect on protoplanetary disk chemistry merits further study. We increased the planetesimal accretion rate in the disk from the nominal value needed to achieve a disk mass comparable to the MMSN to three times that value. Figure \ref{AccretionRate} shows the effect of this increase on the carbon abundance of planetesimals in a disk with an original C/O ratio of 0.8. Higher accretion rates lead to faster depletion of oxygen and a correspondingly fast increase in the local C/O ratio. As a consequence, carbon condenses out of the disk gas at more distant radii leading to carbon enriched planetesimals throughout more of the disk.
        % again use "figure*" instead of "figure" to span both columns

\subsection{N-Body Simulations}
\begin{figure}      % use "figure*" instead of "figure" if you want your figure to span both columns
\epsscale{1.1}      % adjust this number to change the size of your figure
\plotone{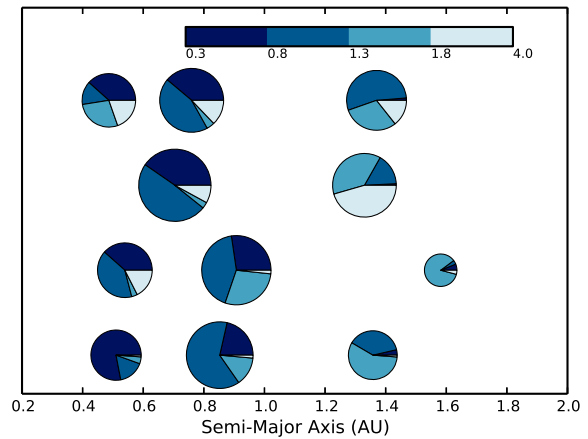}
\caption{\red{ Material source regions for simulated planets. Each pie plot corresponds to a simulated planet and each row to one simulation. Each planet's final orbital position is indicated by its locations along the x-axis and its mass is proportional to the radius of the pie plot cubed. Each colored slice represents material that originated from the region of the disk indicated in the color bar. }}
\label{MaterialOrigin}
\end{figure}

In each of the four N-body simulations, we form two or three planets within 2 AU (see Figure \ref{MaterialOrigin}). They range in mass from 0.1-1.3M$_{\oplus}$ with a median mass of 0.7M$_{\oplus}$. These results are consistent with the CJS simulations of \citet{Obrien06} from which they were based off of. \red{ Figure \ref{MaterialOrigin}  highlights both the importance and stochasticity of radial mixing of planetesimals in the final stages of terrestrial planet formation. The origin of material for any given planet can range from a small area around the planet's final location to practically the whole terrestrial planet forming zone. This stochasticity acts to increase the diversity of exoplanet compositions and also means that any one system is not representative of the others. The main importance of radial mixing is that it weakens any compositional gradients that existed at earlier stages.
}

        % again use "figure*" instead of "figure" to span both columns

\subsection{Planet Bulk Composition}

\begin{figure}      % use "figure*" instead of "figure" if you want your figure to span both columns
\epsscale{1.2}      % adjust this number to change the size of your figure
\plotone{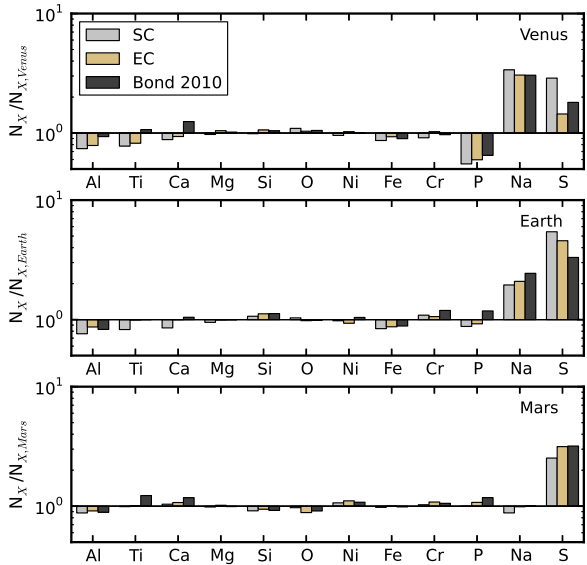}
\caption{Elemental abundances of simulated planets normalized to abundances of corresponding Solar System planets. Results from our sequential condensation (SC) model  and our equilibrium chemistry (EC) model are shown in addition to those of \citet{Bond10}}
\label{SSAbundances}
\end{figure}

We first performed a benchmark test to see if our equilibrium chemistry simulations could reproduce the bulk compositions of the terrestrial planets of the Solar System. \red{As can be seen from Figure \ref{SSAbundances}, the simulated compositions are in good agreement with the Solar System values except for some of the more volatile elements, which are over-abundant in the simulated planets. It is possible that non-equilibrium effects, such as evaporation during planetesimal collisions \citep{Bond10}, were important in the Solar Nebula.} The elemental abundances for the Solar System planets were obtained from \citet{Morgan80}, \citet{Kargel93} and \citet{Lodders97} and are expected to have errors as high as 25\% \citep{Bond10}. Our results are consistent with the findings of \citet{Bond10}, which can also be seen in Figure \ref{SSAbundances}. Small differences exist between our model and theirs, but these are expected due to the stochastic nature of planet formation and differences in the protoplanetary disk model used.

Also shown in Figure \ref{SSAbundances} are the bulk elemental abundances of simulated planets using the sequential condensation chemistry model. In general the results are very similar to those of the equilibrium chemistry. The only elements where we expected differences could occur based on the abundance curves in Figure \ref{DiskComposition} are O, Na and S. These elements have intermediate condensation temperatures and therefore are not in the same proportions throughout the terrestrial planet formation zone. The differences seen in the abundance curves between the sequential condensation and equilibrium chemistry models were likely washed out due to the radial mixing of planetesimals during late stage formation simulations leading to the similarity in bulk compositions between the models.

        % again use "figure*" instead of "figure" to span both columns

\begin{figure}      % use "figure*" instead of "figure" if you want your figure to span both columns
\epsscale{1.2}      % adjust this number to change the size of your figure
\plotone{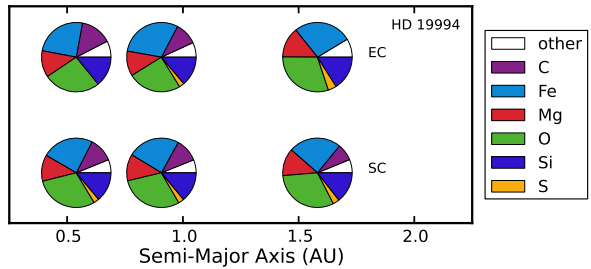}
\caption{Composition of simulated planets in the HD19994 system. Each pie chart shows the fraction of a planet's mass in each of the most abundant elements and its semi-major axis (along the x-axis).  Results for the equilibrium chemistry model are shown in the top row and those for the sequential condensation model are shown in the bottom row.}
\label{HighCarbonAbundances}
\end{figure}

In the simulations of carbon rich systems there are significant differences between the equilibrium chemistry and sequential condensation models. Figure \ref{HighCarbonAbundances} shows the weight percent of the most abundant elements in the simulated planets for the HD 19994 system. The innermost two planets accreted most of their material from a region where carbon was solid in chemical equilibrium at 1.5$\times10^5$ years and thus have a large fraction of their mass in carbon. This results in planets with similar bulk composition between the two models. However, the outermost planet of this system accreted little of its material from the carbon rich region of the disk in the equilibrium chemistry model and consequently has a very small fraction of its mass in carbon. On the other hand, carbon is abundant in planetesimals beyond 2 AU in the sequential condensation model, thus forming a planet with a significant portion of carbon at 1.6AU.

        % again use "figure*" instead of "figure" to span both columns

%%%%%%%%%%%%%%%%%%%%%%%%%%%%%%%%%%%%%%%%%%%%%%%%%%%%%%%%%%%%%%%

\section{Discussion}
 \subsection{Why Should the Sequential Condensation Model be Used?}
 
 	This and previous works both found that an equilibrium chemistry model of the Solar Nebula produces planets with remarkably similar abundances to the true terrestrial planet abundances. The sequential condensation model produces similar results but at the cost of a much more complex model. This begs the question: why use the sequential condensation model at all?
	
	Firstly, the equilibrium chemistry model is not physically motivated in regards to its connection with planetesimal formation. The implicit assumption of the equilibrium chemistry model is that all planetesimals formed at the same time everywhere in the disk and on timescales much shorter than the timescales for disk evolution. For any given region of the disk, planetesimals should begin forming as soon as conditions are right. These conditions should occur at different times for different parts of the disk. Assuming they all form simultaneously throughout the disk is unrealistic.  Furthermore, we know from age dating of meteoritic material that planetesimal formation lasted for about 2.5 Myr \citep{Amelin02}. The assumptions of the equilibrium chemistry model have been overlooked in the past because of the model's success in predicting the bulk composition of the Solar System terrestrial planets.
	
	The success of the equilibrium model can largely be attributed to the high condensation temperatures of most of the major rock forming elements. For all but the hottest disks, these elements will be fully condensed out, and thus in solar proportions, throughout the entire terrestrial planet forming region. The refractory elements are of approximately solar proportions in the terrestrial planets (excluding Mercury) and so the models naturally match these. All that's really required then is a disk model that can reproduce the abundances of the more volatile elements (e.g. O, Na, S, etc). \citet{Bond10} were able to find a model that produced planets with the correct O abundance, but these planets were over abundant in Na and S compared to the Solar System planets. Nevertheless, their model accurately reproduced the abundances of ten out of twelve of the most common elements in the terrestrial planets. 
	
	The situation is different for carbon rich systems. In chemical equilibrium, carbon is abundant as a solid at temperatures higher than $\sim$550 K but is a gas at lower temperatures. For most disk ages, 550 K lies in the middle of the terrestrial planet forming region. Consequently, the predicted composition of planets will vary greatly in the equilibrium chemistry model depending on the choice of disk age. Because there are no carbon rich systems in which we know the composition of the planets, it is not obvious which, if any, disk age will produce the correct results.

	We suggest that the equilibrium chemistry model's ability to reproduce the abundances in the Solar System is a natural consequence of the condensation properties of a disk of solar composition, but that it does not necessarily produce realistic results for carbon rich systems. The sequential condensation model provides a more realistic picture of planetesimal formation in which planetesimals can form in any region of the disk at any time as long as the conditions are correct. This is consistent with the idea that planetesimals form over a few million years, and the model produces planets with elemental abundances in good agreement with those of the Solar System's planets. Although the results for both models are similar in the case of the Solar System, the sequential condensation model likely predicts more realistic compositions for planets in carbon rich systems.

\subsection{Formation and Growth of Planetesimals}
	The most uncertain part of the sequential condensation model is the prescription for planetesimal formation and growth. This is necessarily so due to the current uncertainties in planetesimal formation theory. The prescription used in this work is likely oversimplified. 
	
	We showed that simply increasing the magnitude of the planetesimal accretion rate has a significant effect on the chemistry. Much larger effects would likely occur by changing its functional form. For example, the planetesimal formation rate could be a stronger function of the density of solids in the disk. In such a model, comparatively more material would form planetesimals earlier in the disk's lifetime, when it is denser and hotter, causing the planetesimal disk to be more enhanced in refractory elements and depleted in volatile elements. Such a scenario might be able to reduce the over abundance of the more volatile elements we see in our current Solar System model. Applied to moderately carbon rich disks, this could lead to enough depletion of oxygen early on that solids containing carbon could condense farther away from the star thus increasing the amount of solid carbon in the terrestrial planet formation zone.
	
\subsection{Migration of Planetesimals}
	In the sequential condensation model, planetesimals form at a given distance from the star and remain there for the duration of the simulation. However, planetesimals are expected to migrate radially due to gas drag, and should accrete material from different parts of the disk. Including planetesimal migration is beyond the scope of this work as it would require a knowledge of the distribution of planetesimal sizes and masses. However, it is possible that migration may significantly effect the composition of forming planetesimals and merits future study.

\subsection{C/O Enhancements}
	The sequestration of oxygen in planetesimals has been used as a mechanism to explain the high observed C/O ratios in some gas giant exoplanets \citep{Oberg11, Helling14}. Additionally, the sequestration of oxygen in the outer disk and resulting depletion in the inner disk has been studied in the context of the distribution of water in the Solar Nebula \citep{Ciesla06}, but its influence on the abundances of other chemical species was not considered. \citet{Najita11} pointed out that this depletion may influence the abundances of hydrocarbons in the inner disk and that these differences should be observable. \citet{Carr11} and \citet{Najita11} find a correlation between the HCN/H$_2$O flux ratio and disk masses for T Tauri stars. They suggest that this trend could be explained if a high HCN/H$_2$O flux results from a high C/O ratio and that the C/O ratio is enhanced more in larger disks due to more efficient planetesimal formation. Similarly, \citet{Pascucci13} find higher average HCN/H$_2$O fluxes in brown dwarf disks than T Tauri stars and attribute this to more efficient planetesimal formation in disks around brown dwarfs.  They go on to point out that the C/O ratios needed to explain their observations would have important implications for the compositions of rocky planets forming in these carbon rich regions of the disk.

	Our models explore exactly this point. Indeed, we find that a portion of the disk can contain a substantial amount of solid carbon for disks with an initial C/O ratio as low as 0.65. This finding is unique compared to the equilibrium chemistry model in which carbon rich planetesimals can only form in disks with initial C/O ratios above 0.8. 
	
	The carbon rich region in the sequential condensation model is generally confined to less than $\sim$0.5-0.6 AU for disks with C/O ratios between 0.65 and 0.8. As can be seen from Figure \ref{MaterialOrigin}, this region accounts for a small fraction of a planet's material source region, creating planets with at most a few percent of their mass in carbon. The dynamical simulations in this work are limited to Solar System like initial conditions (i.e., the planetesimal disk extends from 0.3-4 AU). However, it is quite possible that planets in other systems could form much closer to their host stars \citep[e.g.][]{Bond10b, Raymond08b}, and thus potentially accrete more carbon rich material.  \citet{Chiang13} discuss the possibility that most of the known close-in exoplanets formed in situ rather than migrating there. If this is the case or even partially the case, then we predict that carbon rich planets may be significantly more common than previously indicated. 
	
	It is also possible that a further exploration of parameter space in our model will change the radial range where carbon rich planetesimals can exist. There are a number of  factors that could influence where and when oxygen rich material is removed from the disk and turned into planetesimals, including: 
	(1) The magnitude and functional form of the planetesimal accretion rate. We showed that by increasing the overall planetesimal accretion rate, the carbon rich region of the disk moved inwards. Additionally, the rate at which planetesimals in different parts of the disk form and grow may not be governed by equation \ref{planetesimalformationrate}. Variation from this form would completely change the pattern of oxygen depletion and thus the C/O enhancement. 
	(2) The disk temperature and density structure as a function of time. The structure and evolution of the disk is undoubtedly connected to the formation and growth of planetesimals. In our planetesimal formation prescription this is true because the planetesimal formation rate is a dependent on the surface density of the disk. By changing the equations of disk evolution we would also be changing the timing of planetesimal formation and thus their composition as well.
	 (3) The overall amount of (C+O)/(Mg+Si). For a fixed C/O ratio, if oxygen (and carbon) are depleted relative to their solar values but Mg and Si are not, then the relative fraction of O that bonds with Mg and Si is higher and thus the effective O depletion is larger and consequently the C/O enhancement is stronger.

%%%%%%%%%%%%%%%%%%%%%%%%%%%%%%%%%%%%%%%%%%%%%%%%%%%%%%%%%%%%%%%

\section{Summary and Conclusions}
We have implemented a model of protoplanetary disk chemistry that accounts for the time evolution of the disk and the formation and accretion of planetesimals. Previous models performed equilibrium chemistry calculations at a single point in the disk's lifetime. We have compared the results of the sequential condensation model to those of the equilibrium chemistry model and find three main results:
\begin{itemize}

\item For a disk of solar composition, the sequential condensation model produces planets with elemental abundances very similar to those produced by the equilibrium chemistry model. Of the elements considered, the abundances match the true values of the Solar System terrestrial planets remarkably well with the exception of sodium and sulfur. Whereas the equilibrium chemistry model was already optimized to reproduce the compositions of the Solar System planets, the full parameter space of the sequential condensation model has yet to be explored. Tuning the sequential condensation model should produce better fits to the abundances of the more volatile elements in the Solar System planets.

\item In carbon rich disks, the sequential condensation model predicts planetesimals rich in carbon over a wider range of orbital radii than the equilibrium chemistry model. When compared to the Solar System optimized case, this results in carbon enriched planets at larger semi-major axes. Overall, the sequential condensation model produces planetesimals with compositions that vary less over the radial range of the terrestrial planet forming region.

\item The sequential condensation model predicts that carbon enriched planetesimals can form in initially oxygen dominated protoplanetary disks. These planetesimals are only produced in the innermost regions of the disk using the current model parameters. However, it is possible that an exploration of parameter space may find that the sequential condensation model can produce carbon rich planetesimals at greater distances from the star. Additionally, if a significant fraction of known close in super-Earths formed in situ, the number of carbon enriched exoplanets may be significantly more than previously indicated.  

\end{itemize}

\acknowledgments 
We thank the referee, John Chambers, for helpful comments used to improved this paper. NM acknowledges support from the Yale Center for Astronomy and Astrophysics (YCAA) at Yale University through the YCAA postdoctoral prize fellowship.

\bibliography{mybib}

\end{document}